\documentclass[12pt,letterpaper]{article}

\usepackage{amsmath}
\usepackage{amssymb}
\usepackage{graphicx}
\usepackage{float}     
\usepackage[titletoc]{appendix}
\usepackage[hidelinks]{hyperref}  
\usepackage{authblk}  
\usepackage{lineno}   
\usepackage[export]{adjustbox}
\usepackage{caption}
\usepackage{subcaption}

\newcommand{\tx}[1]{\text{#1}}

\begin{document}
\pagestyle{plain}

\pagestyle{plain}

\title{
The Position Resolution of a Charged Particle Hodoscope Consisting of Triangular 
Scintillator Counters
}
\author{E. Craig Dukes}
\affil{Physics Department, University of Virginia, Charlottesville, VA, 22904, USA}
\date{October 27, 2026}

\maketitle

\begin{abstract}
Hodoscopes consisting of triangular counters with embedded wavelength shifting fibers
are becoming more popular as they are relatively inexpensive, are easily fabricated,
need no cooling systems, employ low-cost photodetectors, require only modest voltages,
and perhaps most importantly, have good position resolution.
Through interpolation of the light yields between adjacent counters, position resolutions
that are typically an order of magnitude better than the counter separation can be obtained.
Here, after a brief explanation of how the interpolation is done to improve the position
resolution, we derive an approximate expressions for the uncertainty in the interpolated
$x$ and $y$ positions as a function of the impact position of the charged particle.
\end{abstract}

\section{Introduction}
\label{intro}
Hodoscopes consisting of arrays of triangular scintillator counters are becoming increasingly
popular due to their modest cost, ease of fabrication, ease of operation, and their ability
to operate in harsh environments. More importantly, position resolutions better than one-tenth of the distance
between the triangle centers have been achieved (see Refs.\,\cite{dzero,muray,minerva,nessie,mima,icarus,lumis},
for example).    We first derive the simple expressions that give the $x$ and $y$ coordinates of a charged particle 
going through a plane of triangular scintillator counters by interpolation of the charged 
particle path lengths through the adjacent counters using the relative light yields in each counter.  
We then derive an approximate expression of the uncertainties in the positions using Poisson estimates
for the light yield distributions.

\section{Determining the $\boldsymbol{x}$ and $\boldsymbol{y}$ Coordinates}
\label{sec:coordinates}

We derive below the $x$ and $y$ positions found by interpolation of the light yields in 
the adjacent counters shown in Fig.\,\ref{fig:one} and Fig.\,\ref{fig:two} below.
The results for the interpolated $x$- and $y$-coordinates are:
\begin{align}
x_{\pm} & = \pm\frac{w}{4}\left(\frac{E_1-E_2}{E_1 + E_2}\right), \label{eq:x} \\
y       & = \frac{h}{2}\left(\frac{E_1 - E_2}{E_1 + E_2}\right),  \label{eq:y}
\end{align}
where the $+$ sign in Eq.\,\ref{eq:x} refers to the case shown in Fig.\,\ref{fig:one}
where the leftmost triangle is inverted and the $-$ sign in Eq.\,\ref{eq:x} refers
to the case shown in Fig.\,\ref{fig:two} where the leftmost triangle is upright.
In both of the above equations $w$ is the triangle base width, $h$ is the triangle 
height, and $E_1$ and $E_1$ are the charged particle path lengths shown in Figs.\,\ref{fig:one}
and \ref{fig:two}. 

To experimentally determine the $x$ and $y$ positions, we replace the path lengths
$E_1$ and $E_2$ with their photoelectron yield proxies, $N_1$ and $N_2$, to get:
\begin{align}
x_{\pm} & \approx \pm\frac{w}{4}\left(\frac{N_1-N_2}{N_1 + N_2}\right), \label{eq:xa} \\
y       & \approx \frac{h}{2}\left(\frac{N_1 - N_2}{N_1 + N_2}\right),  \label{eq:ya}
\end{align}
where the approximate sign indicates the fact that $N_1$ and $N_2$ are imperfect measures
of path lengths $E_1$ and $E_2$.

\subsection*{Case 1: Charged Particle Traverses the Junction Between the Left and Center Counters}

We derive Eq.\,\ref{eq:x} and Eq.\,\ref{eq:y} referring to the geometry shown in Fig.\,\ref{fig:one} below. 
Each counter is an isosceles triangle of width $w$ and height $h$.
We assume that the charged particle's incident angle is such that it only traverses no more than two counters,
and we assume that the two counters have a perfect isosceles triangular shape.
The origin of the coordinate system is at the midpoint of the interface between the two counters:
$\boldsymbol{M}_{+}$ in  Fig.\,\ref{fig:one}.
The coordinates of the point at which a charged particle passes the boundary between two counters 
is $\boldsymbol{P}_{+}$ in  Fig.\,\ref{fig:one}.
The charged particle's path (shown in blue in Fig.~\ref{fig:one}), has a length $E_2$ in the left
(inverted) counter and $E_1$ in the center (upright) counter.

\begin{figure}[htb]
\centerline{\includegraphics[width=5.5in]{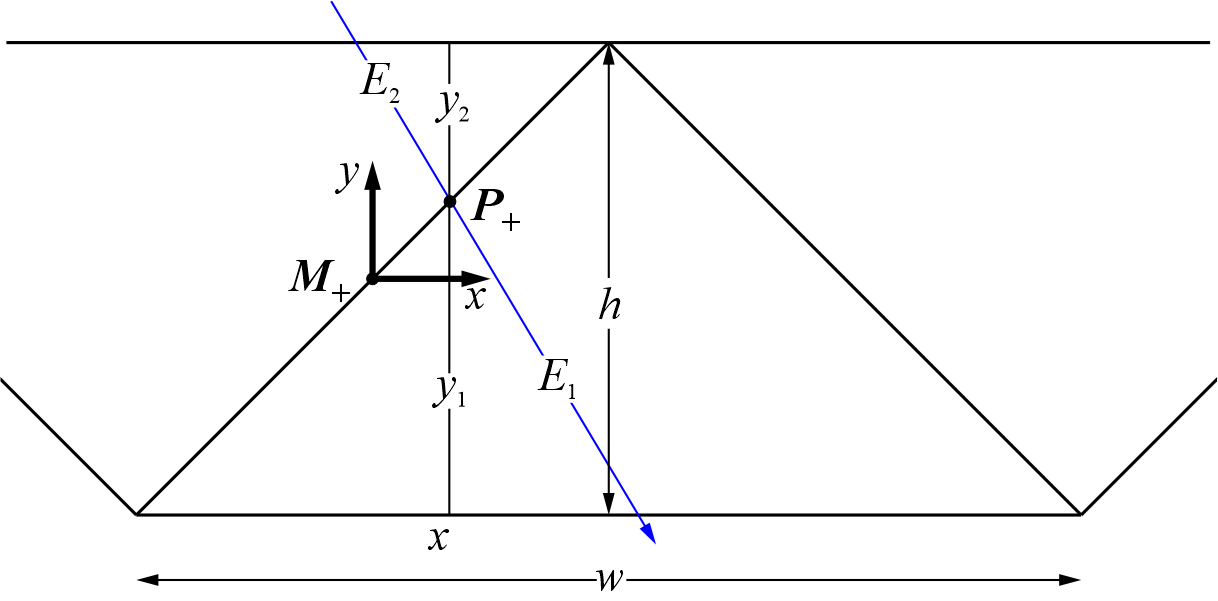}}
\caption{The counter geometry showing a charged track (in blue) traversing the inverted counter
to the left of the center (upright) counter.}
\label{fig:one}
\end{figure}

We wish to find the best estimate of the coordinates of the point $\boldsymbol{P_+}$($x$, $y$),
relative to the origin of the coordinate system  $\boldsymbol{M}_{+}$ in Fig.~\ref{fig:one}.
We have from the geometry of similar triangles:
\begin{equation*}
\frac{y_1}{y_2} = \frac{E_1}{E_2} \tx{~~~~or~~~~} y_2 = \frac{E_2}{E_1}y_1.
\label{eq:ratio-left}
\end{equation*}
In addition, $y_1 + y_2$ equals the height of the counter, or:
\begin{equation*}
y_2 = h - y_1.
\end{equation*}
Setting these two above  expressions for $y_2$ equal to one another, we have:
\begin{equation*}
y_1 = h\frac{E_1}{E_1 + E_2}.
\end{equation*}
We can now solve for $y$:
\begin{align}
y & = y_1 - \frac{h}{2} = h\frac{E_1}{E_1 + E_2} - \frac{h}{2}, \notag \\
y & = \frac{h}{2}\left(\frac{E_1 - E_2}{E_1 + E_2}\right).
\end{align}

Does this equation make sense?  If the path lengths $E_1$ and $E_2$ are equal,
then $y = 0$, which is correct.  If $E_1 = 0$ then $y = -h/2$, and if $E_2 = 0$, 
then $y = +h/2$, both of which are correct.

Once the $y$-coordinate is known ($y_1$), the $x$-coordinate can be found using
the slope: 
\begin{equation*}
S = \frac{\Delta y}{\Delta x} = \frac{h}{w/2} = \frac{2h}{w}. 
\end{equation*}
The $x$-coordinate is then:
\begin{align}
x & = \frac{y}{S} = \frac{w}{2h}\frac{h}{2}\left(\frac{E_1 - E_2}{E_1 + E_2}\right), \notag \\
x & = \frac{w}{4}\left(\frac{E_1-E_2}{E_1 + E_2}\right).
\end{align}

Does this equation make sense?  If the path lengths $E_1$ and $E_2$ are equal,
then $x = 0$, which is correct.  If $E_1 = 0$ then $x = -w/4$, and if 
$E_2 = 0$, then $x = +w/4$, both of which are correct.

\subsection*{Case 2: The Charged Particle Traverses the Junction Between the Center and Right Counters}

\begin{figure}[htb]
\centerline{\includegraphics[width=5.5in]{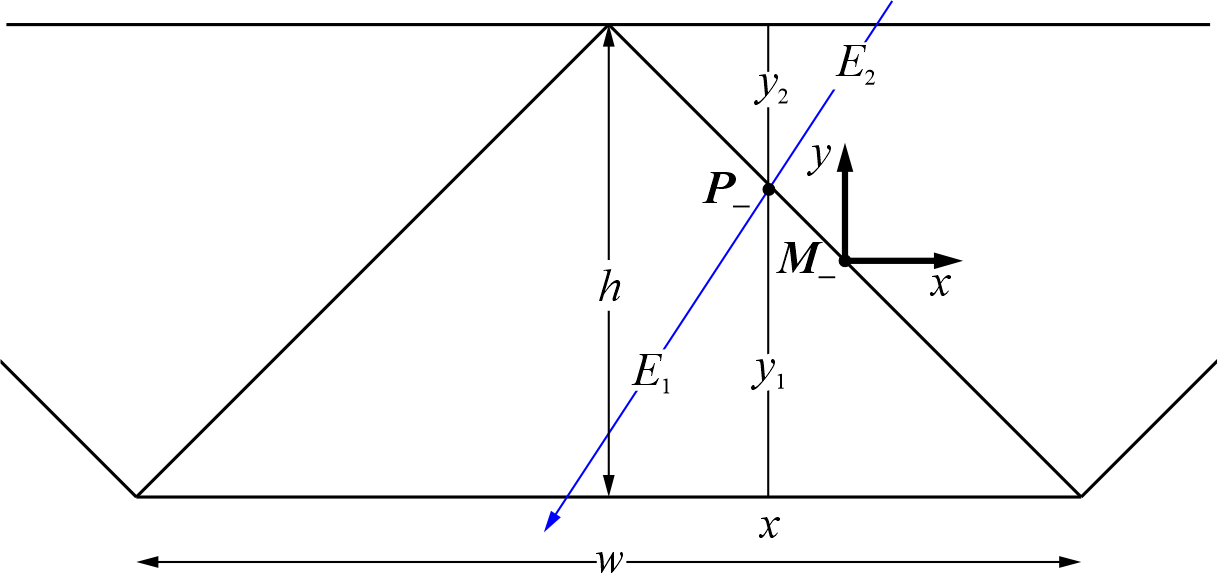}}
\caption{The counter geometry showing a charged track (in blue) traversing the inverted counter
to the right of the center (upright) counter.}
\label{fig:two}
\end{figure}

This derivation proceeds exactly as that shown above for Case\,1, and hence will not be
repeated. The only difference is that the in going from the $y$ coordinate to the $x$ coordinate
a minus sign appears as the slope of the junction between the two counters, 
as shown in Fig.\,\ref{fig:two} is negative.

We have for the $y$ position in Fig.\,\ref{fig:two}:
\begin{equation}
y  = \frac{h}{2}\left(\frac{E_1 - E_2}{E_1 + E_2}\right),
\end{equation}
exactly the same equation as found in Case\,1 above.

Does this equation make sense?  If the path lengths $E_1$ and $E_2$ are equal,
then $y = 0$, which is correct.  
If $E_1 = 0$ then $y = -h/2$, and if $E_2 = 0$, then $y = +h/2$, both of which are correct.

The $x$-coordinate is:
\begin{equation}
x  = -\frac{w}{4}\left(\frac{E_1-E_2}{E_1 + E_2}\right).
\label{eq:xm}
\end{equation}
The minus sign in front of Eq.\,\ref{eq:xm} is the only difference between 
this value for $x$ and that from the previous section.

Does this equation make sense?  If the path lengths $E_1$ and $E_2$ are equal,
then $x = 0$, which is correct.   If $E_1 = 0$, then $x = +w/4$,
and if $E_2 = 0$, then $x = -w/4$, both of which are correct.

\subsection*{Angular Determination}

\begin{figure}[htbp]
\centerline{\includegraphics[width=4.0in]{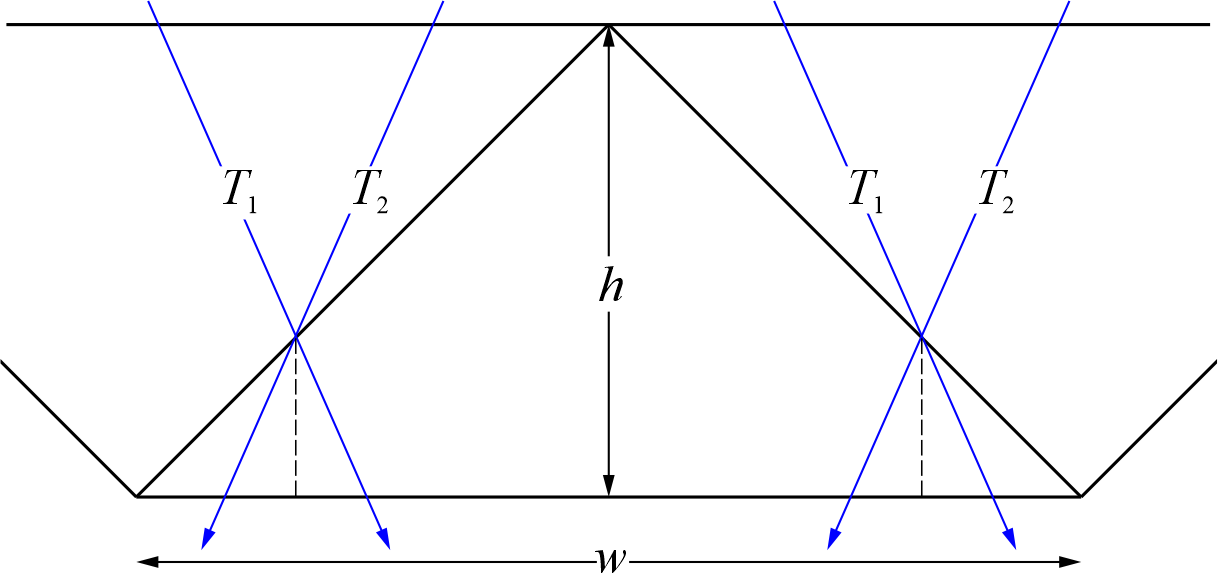}}
\caption{Two pairs of charged particles going through adjacent triangular counters.
Each track of each pair has the same $x$ and $y$ coordinates, although their
incident angles are quite different.}
\label{fig:angular}
\end{figure}

We emphasize that the derivations of Eq.\,\ref{eq:x} and Eq.\,\ref{eq:y} are independent
of the charged particle incident angle, {\em as long as it does not traverse more than 
two counters}.  And, indeed, the incident angle cannot be determined by this method.
Figure\,\ref{fig:angular} shows two charged-particle tracks, $T_1$ and $T_2$ going through the left and
right counter junctions at exactly the same $x$ and $y$ points, but at widely different angles.
The point of this is to show that one cannot tell the angle from only one plane of triangular 
counters.  Another plane is needed.

%
%

\section{The Interpolated Position Uncertainty}

We replace the distances traversed by the charged particles in the counters, $E_1$ and $E_2$
by their photoelectron yield proxies, $N_1$ and $N_2$, respectively in Eq.\,\ref{eq:x}.
We have then
\begin{align}
x & = \pm\frac{w}{4}\left(\frac{N_1 - N_2}{N_1 + N_2}\right). \notag
\end{align}
The uncertainty in $x$ as a function of the uncertainty in $N_1$ and $N_2$:
\begin{alignat}{2}
\frac{\partial x}{\partial N_1} & =  \pm\frac{w}{4}\frac{(N_1 + N_2) - (N_1-N_2)}{(N_1 + N_2)^2} 
                               && =  \pm\frac{w}{2}\frac{N_2}{(N_1 + N_2)^2}, \label{eq:part1} \\
\frac{\partial x}{\partial N_2} & =  \pm\frac{w}{4}\frac{-(N_1+N_2)-(N_1 - N_2)}{(N_1 + N_2)^2}
                               && =  \pm\frac{w}{2}\frac{-N_1}{(N_1 + N_2)^2}. \label{eq:part2} 
\end{alignat}
We add the two uncertainties in quadrature to get the total uncertainty in $x$:
\begin{equation}
(\Delta x)^2 = \left(\frac{\partial x}{\partial N_1}\right)^2(\Delta N_1)^2 + 
               \left(\frac{\partial x}{\partial N_2}\right)^2(\Delta N_2)^2.
\label{eq:errory}
\end{equation}
We wish to estimate the uncertainty in parameters $N_1$ and $N_2$. For charged particles going
though a thin counter, this is given by a Landau distribution \cite{landau}. However, to simplify our derivation
we assume the light yield follows a Poisson distribution in order to allow a simple estimation of their uncertainties: 
\[ 
(\Delta N_1)^2 = \sigma_1^2 = N_1 \textrm{~~~and~~~} (\Delta N_2)^2 = \sigma_2^2 = N_2.
\]
Using these two values for $\sigma_1$ and $\sigma_2$ we find, using Eq.\,\ref{eq:errory} in conjunction with
Eqs.\,\ref{eq:part1} and \ref{eq:part2}:
\begin{align}
\sigma_x^2 & = \left(\frac{w}{2}\right)^2\left[\frac{N^2_2}{(N_1 + N_2)^4}N_1 + \frac{N^2_1}{(N_1 + N_2)^4}N_2\right], \notag \\
             & = \left(\frac{w}{2}\right)^2\frac{N_1N_2}{(N_1 + N_2)^4}(N_1 + N_2), \notag \\
             & = \left(\frac{w}{2}\right)^2\frac{N_1N_2}{(N_1 + N_2)^3}. \notag
\end{align}
Finally,
\begin{equation}
\sigma_x = \left(\frac{w}{2}\right)\sqrt{\frac{N_1 N_2}{(N_1 + N_2)^3}}.
\label{eq:delx}
\end{equation}
The $y$ position uncertainty, $\sigma_y$, is easily determined to be:
\begin{equation}
\sigma_y  = h\sqrt{\frac{N_1N_2}{(N_1 + N_2)^3}}.
\label{eq:dely}
\end{equation}
Note that Eq.\,\ref{eq:delx} and Eq.\,\ref{eq:dely} are true for both Case\,1 and Case\,2 described above.

Do these equations make sense? If $N_1$ or $N_2$ is $0$ then the charged particle is going
through the apex of one of the counters and $\sigma_x$ or $\sigma_y$ is zero.

At what value of $x$ is the uncertainty maximal?
To find out we set the derivative of Eq.~\ref{eq:delx} equal to zero.
We have, setting $N_2 = N_T - N_1$, where $N_T = N_1 + N_2$:
\begin{align}
\frac{\partial \sigma_x}{\partial N_1} 
  & =  \left(\frac{w}{2}\right)\frac{\partial}{\partial N_1}\left(\frac{\sqrt{N_1(N_T - N_1)}}{N_T^{3/2}}\right), \notag \\
  & = \left(\frac{w}{2}\right)\frac{1}{2N_T^{3/2}}\frac{(N_T-N_1) - N_1}{\sqrt{N_1(N_T-N_1)}}, \notag \\
\frac{\partial \sigma_x}{\partial N_1}
  & = \frac{w}{4N_T^{3/2}}\frac{N_2 - N_1}{\sqrt{N_1N_2}}.
\end{align}
We also have a similar expression for the derivative of the $y$ position uncertainty:
\begin{equation}
\frac{\partial \sigma_y}{\partial N_1}
   = \frac{h}{2N_T^{3/2}}\frac{N_2 - N_1}{\sqrt{N_1N_2}}.
\label{eq:maxy}
\end{equation}

Both equations are zero when $N_1 = N_2$, which happens when the charged particle crosses 
midway up the junction between the two counters where the path lengths through the adjacent
counters are equal.  If we let $N_1 = N_2 = \tfrac{1}{2}N_T$, then Eq.\,\ref{eq:delx} 
reads:
\begin{equation}
\sigma_{x\tx{ max}} = \left(\frac{w}{2}\right)\sqrt{\frac{\tfrac{1}{4}N_T N_T}{(N_T)^3}}
         = \left(\frac{w}{4}\right)\sqrt{\frac{1}{N_T}} .
\end{equation}
\begin{figure}[htbp]
\centerline{\includegraphics[width=4.5in]{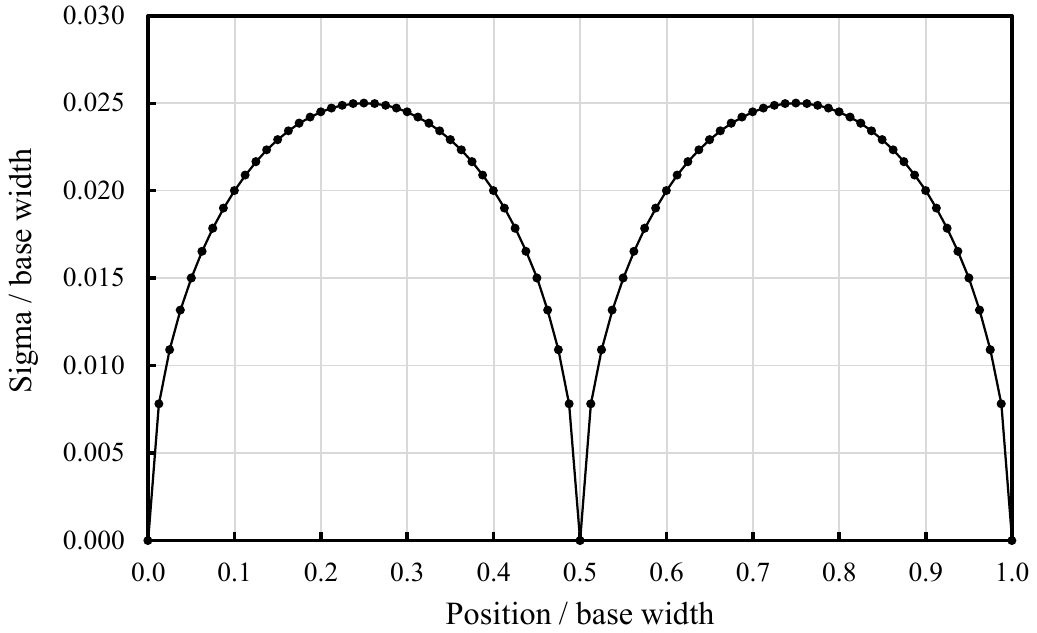}}
\captionsetup{width=5.5in}
\caption{The $x$ position uncertainty for normally incident particles versus the $x$ position 
along the base of the center triangular counter, from Eq.\,\ref{eq:delx}. 
Units are counter base width ($w$) and a light yield of 100\,PE/$h$ is assumed. The resolution is 
best at the corners of the center triangular counter and worst midway up its two sides.}
\label{fig:unitless}
\end{figure}

\begin{figure}[htbp]
\centerline{\includegraphics[width=4.5in]{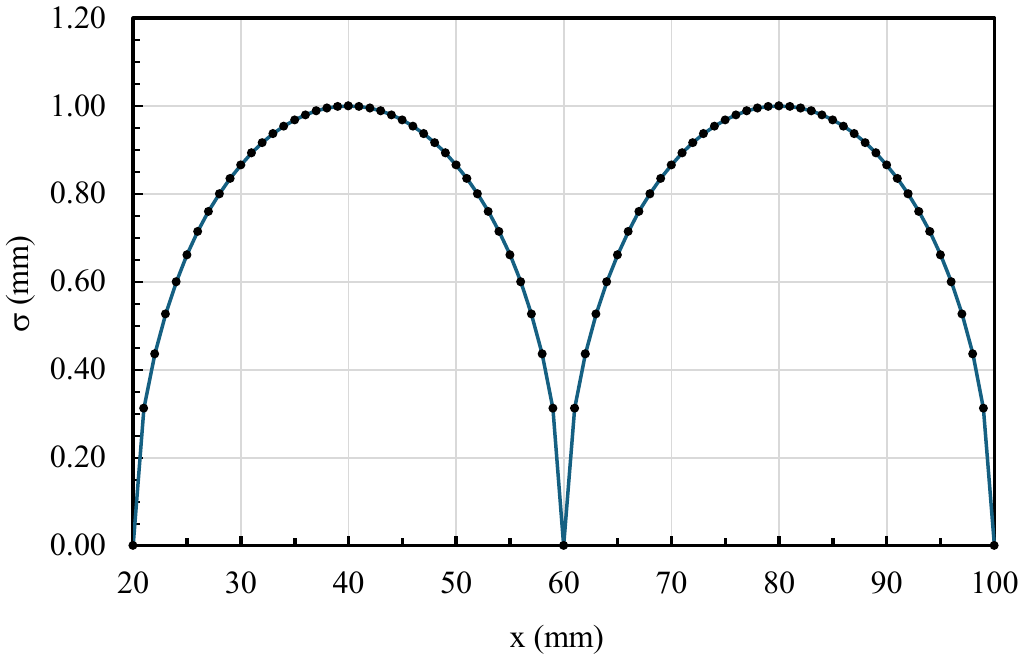}}
\captionsetup{width=5.5in}
\caption{The $x$ position uncertainty for normally incident particles versus the $x$ position 
along the base of the center triangular counter, from Eq.\,\ref{eq:delx}. The counter is 
$40 \times 20$\,mm$^2$ ($w \times h$) and a light yield of 5\,PE/mm is assumed. The resolution is 
best at the corners of the center triangular counter and worst midway up its two sides.}
\label{fig:real}
\end{figure}

In Fig.~\ref{fig:unitless} the $x$-position uncertainty is plotted as a function of the $x$ position in 
units of the base width, $w$, assuming a photoelectron yield of 100\,PE/$h$ (a light yield
achieved by Ref.\,\cite{triangles}). Figure\,\ref{fig:real} shows the resolution for a counter of 
$40 \times 20$\,mm$^2$ ($w \times h$) and the same PE yield.
The greatest uncertainty occurs when the charged particle crosses halfway up the junction between adjacent
counters, and has a value of $\sigma_x = 0.025{\cdot}w$ for a photoelectron
yield of 5\,PE/mm.
The position resolution averaged over all incident $x$ values is $\sigma_{x \tx{ av}} = 0.019{\cdot}w$.
For a triangular counter of width 40\,mm, such as the counter described in Ref.\,\cite{triangles},
that resolution is $\sigma_{x \tx{ av}} = 0.76$\,mm,
using their measured photoelectron yield of 100\,PE/$h$.  The worse resolution for such a counter is 1.0\,mm.

Our result is a lower limit on the position uncertainty one expects from a real hodoscope.
Real position resolutions are worse for several reasons, such as imperfections in the counter geometry,
electronics noise, and the fact that assuming a Poisson
distribution for the photoelectron yield underestimates the tails in the real light yield
distributions.  A recent study of the position resolution of a hodoscope consisting of  $40 \times 20$\,mm$^2$
($w \times h$) counters with a light yield of 100\,PE/$h$ (and using Eq.\,\ref{eq:x} to
determine the $x$ position) shows a similar behavior in the position resolution as a function of
the charged particle impact position as shown in Fig.\,\ref{fig:real} \cite{triangles}.
They find a maximum $x$-position uncertainty of roughly 2\,mm, about a factor of two worse than 
what we estimate here.  However, with a toy Monte Carlo that uses a Landau$+$Gaussian fit to
the measured light yield distribution, they find excellent agreement with the data.


\end{document}